# Luminescent Platform for Thermal Sensing and Imaging Based on Structural Phase-Transition

Anam Javaid[1], Maja Szymczak[1*], Malgorzata Kubicka[1], Vasyl Kinzhybalo[1], Marek Drozd[1], Damian Szymanski[1], Lukasz Marciniak[1*]

[1] Institute of Low Temperature and Structure Research, Polish Academy of Sciences,

Okólna 2, 50-422 Wrocław, Poland

*corresponding author: l.marciniak@intibs.pl

m.szymczak@intibs.pl

**Abstract**

The remarkable sensitivity of the luminescent properties of $Eu^{3+}$ ions to structural changes in host materials has been well-explored for years. However, the application of this feature of $Eu^{3+}$ in materials exhibiting thermally induced structural phase transitions for the development of luminescent thermometers has only recently been proposed. The narrow operating range of such thermometers necessitates the exploration of new host materials. In response to this demand, this study carefully analyzes the spectroscopic properties of $Na_3Sc_2(PO_4)_3$:$Eu^{3+}$ as a function of temperature and dopant ion concentration. As demonstrated, $Na_3Sc_2(PO_4)_3$:$Eu^{3+}$ undergoes a phase transition from a low-temperature monoclinic phase to a high-temperature trigonal structure, resulting in significant changes in both the emission spectrum shape of $Eu^{3+}$ ions and the depopulation kinetics of the $^5D_0$ level. Consequently, $Na_3Sc_2(PO_4)_3$:$Eu^{3+}$ can be utilized as both a ratiometric and a lifetime-based luminescence thermometer, achieving maximal relative sensitivities of 3.4% $K^{-1}$ and 1.0% $K^{-1}$ or the respective approaches. Additionally, this work highlights how increasing the concentration of $Eu^{3+}$ ions enables the tuning of the thermal operating range to achieve optimal thermometric performance. Moreover, an implementation of ratiometric approach of temperature sensing and imaging with $Na_3Sc_2(PO_4)_3$:$Eu^{3+}$ using digital camera without filters was demonstrated. This is the first report that demonstrates thermal imaging using $Eu^{3+}$-solely doped phosphor. This finding underscores the potential of $Na_3Sc_2(PO_4)_3$:$Eu^{3+}$ as a versatile host material for advanced luminescent thermometry applications.

**Introduction**

Luminescent thermometry, a technique for temperature measurement based on changes in the spectroscopic properties of a phosphor, has seen significant advancement in recent years due to its unique capabilities[1–8]. Chief among these are remote, rapid, and dynamic readout[9,10]. Moreover, luminescent thermometry is one of the few techniques that enables not only point-specific temperature readout but also thermal imaging of the analysed object[11–14]. Of the various spectroscopic parameters of the phosphor that can be utilized for temperature measurement, the ratiometric and lifetime-based approaches offer the highest reading reliability. In these approaches, temperature is determined through the thermal change in either the luminescence intensity ratio or luminescence decay[3,15]. However, due to physical constraints, the thermal operating range of luminescent thermometers is limited to the range within which the luminescence intensity of one of the monitored bands remains detectable. Conversely, a dynamic thermal response in the measured parameter favours high relative sensitivity in luminescent thermometers[9,10,16]. Considering these factors, two main directions have emerged in the development of luminescent thermometry: (i) thermometers with relatively low sensitivity but a wide thermal operating range[17–21], and (ii) thermometers with high sensitivity operating within a narrow thermal range [7,16,22–24]. Although both of these directions are equally important, the choice of which of these types of luminescent thermometers should be used is determined by the requirements of a specific application. In many cases, the thermal operating range of a given system or device requiring temperature measurement is relatively not wide and in such a case high relative sensitivity is highly desirable[25–29].

One recent approach to achieving high sensitivity within a narrow thermal range involves thermometers that utilize thermally induced structural phase transitions[30–35]. In this method, the structural phase transition alters the point symmetry of the crystallographic site occupied by the luminescent ion, resulting in a change in the emission spectrum shape and enabling a ratiometric approach to temperature measurement. An important advantage of this approach, aside from high relative sensitivity, is the ability to adjust the operating range by altering the host material composition to suit application requirements[32,33]. However, this tunning is feasible only within a certain thermal range.

Therefore, further research on the implementation of other host materials into phase-transition-based luminescence thermometers is highly desirable. In this work, we present a study on the impact of the thermally induced structural transition in $Na_3Sc_2(PO_4)_3$ on the spectroscopic properties of doped $Eu^{3+}$ ions. As is well known, in $Na_3Sc_2(PO_4)_3$, temperature changes lead to two structural phase transitions, a feature highly relevant for luminescent thermometry[36–47]. The observed thermal variations in the spectroscopic properties of $Na_3Sc_2(PO_4)_3:Eu^{3+}$, manifesting as changes in the emission spectrum shape and depopulation kinetics of the $^5D_0$ level of $Eu^{3+}$ ions, were examined as a function of $Eu^{3+}$ ion concentration. These studies demonstrate the potential of $Na_3Sc_2(PO_4)_3:Eu^{3+}$ for use in both ratiometric and lifetime-based luminescent thermometry.

**Experimental section**

*Synthesis*

A series of powder samples of $Na_3Sc_2(PO_4)_3$:x%$Eu^{3+}$, (where x = 0.1, 0.2, 0.3, 0.4, 0.5, 1, 2, 5) were synthesized using a conventional high-temperature solid-state reaction technique. $Na_2CO_3$ (99.9% of purity, Alfa Aesar), $Sc_2O_3$ (99.9% of purity, Alfa Aesar), $NH_4H_2PO_4$ (99.9% of purity, POL-AURA), and $Eu_2O_3$ (99.999 % of purity, Stanford Materials Corporation)) were used as starting materials. The stoichiometric amounts of reagents were finely ground in an agate mortar with few drops of hexane and, then annealed in the alumina crucibles at 1573 K for 5 h (heating rate of 10 K $min^{-1}$) in air. The final powders were cooled to the room temperature and then ground again.

*Characterization*

The obtained materials were examined using powder X-ray diffraction technique. Powder diffraction data were obtained in Bragg–Brentano geometry using a PANalytical X'Pert Pro diffractometer equipped with Oxford Cryosystems Phenix (low-temperature measurements) attachment using Ni-filtered Cu K$\alpha$ radiation (V=40 kV, I=30 mA). Diffraction patterns in 2θ range of 15-90º were measured in cooling/heating sequence in the temperature range from 320 to 80 K. ICSD database entries No. 56865 (LT phase) and 65407 (HT phase) were taken as initial models for the analysis of the obtained diffraction data.

Morphology and chemical composition of the studied samples were determined with a Field Emission Scanning Electron Microscope (FE-SEM, FEI Nova NanoSEM 230) equipped with an energy-dispersive X-ray spectrometer (EDX, EDAX Apollo X Silicon Drift Correction) compatible with Genesis EDAX microanalysis Software. Before SEM imaging, the $Na_3Sc_2(PO_4)_3$:0.2%$Eu^{3+}$ sample (as a representative sample in the entire study series) was dispersed in alcohol, and then a drop of suspension was placed in the carbon stub. Finally, SEM images were recorded with an accelerating voltage of 5.0 kV in a beam deceleration mode which improves imaging parameters such as resolution and contras as well as reduces contamination. In the case of EDS measurements the sample was scanned at 30 kV.

A differential scanning calorimetric (DSC) measurements were performed using Perkin-Elmer DSC 8000 calorimeter equipped with Controlled Liquid Nitrogen Accessory LN2 with a heating/cooling rate of 20 K min$^{-1}$. The sample was sealed in the aluminum pan. The measurement was performed for the powder sample in the 100 - 800 K temperature range. The excitation and emission spectra were obtained using the FLS1000 Fluorescence Spectrometer from Edinburgh Instruments equipped with 450 W Xenon lamp and R928 photomultiplier tube from Hamamatsu. Luminescence decay profiles were also recorded using the FLS1000 equipped with 150 W µFlash lamp. The average lifetime ($\tau_{avr}$, Eq. 1) of the excited levels was calculated based on fit of the luminescence decay profiles by double-exponential function (Eq. 2):

$$\tau_{avr} = \frac{A_1\tau_1^2 + A_2\tau_2^2}{A_1\tau_1 + A_2\tau_2} \qquad (1)$$

$$I(t) = I_0 + A_1 \cdot \exp\left(-\frac{t}{\tau_1}\right) + A_2 \cdot \exp\left(-\frac{t}{\tau_2}\right) \qquad (2)$$

where $\tau_1$ and $\tau_2$ represent the luminescence decay parameters and $A_1$, $A_2$ are the fitted amplitudes of the double-exponential function. During the temperature-dependent emission measurements, the temperature of the sample was controlled by a THMS600 heating–cooling stage from Linkam (0.1 K temperature stability and 0.1 K set point resolution).

The digital images were taken using a Canon EOS 400D camera with a EFS 60 mm macro lens using a 1 s integration time, 14.3 lp/mm spatial resolution and long pass 450 nm filter from Thorlabs. After capturing color images the emission maps for red and green channels (RGB) were extracted using IrfanView 64 4.51 software. The R and G intensity maps were divided by each other using ImageJ 1.8.0_172 software. This procedure was also used for phosphor placed at constant temperatures to create a calibration curve used for R/G maps into thermal maps conversion. Thermovision images were collected using a T540 camera from FLIR.

**Results and Discussion**

It is widely reported that the $Na_3Sc_2(PO_4)_3$ exists in three crystal structures ($\alpha$, $\beta$ and $\gamma$ phases), depending on the synthesis conditions and ionic substitution schemes [36–45,48]. The $\alpha$ phase is a monoclinic structure with a space group *Bb*, while the $\beta$ and $\gamma$ phases have a trigonal structure with *R3c* space group (Figure 1a)[49]. All three phases possess a structure composed of $ScO_6$ octahedra, $PO_4$ tetrahedra and 5, 7 and 8 - coordinated by $O^{2-}$ $Na^+$ ions. For both phases, the most preferential location for $Eu^{3+}$ dopant ions is the 6-fold coordinated $Sc^{3+}$ ions, due to the similarity in ionic radius as well as the same ionic charge. The room temperature XRD patterns of $Na_3Sc_2(PO_4)_3$ with different concentration of $Eu^{3+}$ correlate well with the reference patterns Figure 1b, see also Figure S1) [49]. However, the Rietveld refinement of the obtained XRD patterns revealed small (~1.2%) amount of $ScPO_4$ (Figure S2-S30). Due to the its small amount and the fact that $ScPO_4$ does not undergo of thermally induced phase transition in the analyzed thermal range its presence does not have influence on the optical properties investigated in this paper.

As reported in the literature, the $Na_3Sc_2(PO_4)$ undergoes two first-order reversible, temperature-induced phase transitions: $\alpha \leftrightarrow \beta$ at T ~ 313–337 K[48] and $\beta \leftrightarrow \gamma$ at ~439 K[50] (during heating). However, any evidence of $\beta \leftrightarrow \gamma$ phase transition was recognized based on the XRD patterns analysis. The analysis of the XRD pattern of $Na_3Sc_2(PO_4):0.2\%Eu^{3+}$ measured as a function of temperature indicates that at room temperature sample consist almost only of the low temperature $\alpha$ phase. An increase in temperature results in gradual increase of the amount of high temperature $\beta$ phase and around 350K its contribution reached 50%. Above 380 K only $\beta$ phase was observed. Due to the smaller ionic radii of $Eu^{3+}$ ions in respect to the host $Sc^{3+}$ cations an increase in the doping level results in a slight decrease of the phase transition temperature. Therefore, the contribution of the $\beta$ phase enhances in room temperature XRD patterns with an increase in $Eu^{3+}$ concentration (Figure 1d). The highest value reaching 16.8% was obtained for $Na_3Sc_2(PO_4):5\%Eu^{3+}$. To trance the influence of $Eu^{3+}$ dopant concentration on $\beta \leftrightarrow \gamma$ phase transition temperature the DSC analysis was performed (Figure 1e). The obtained results revealed that the phase transition temperature monotonically decreases from 340 K for $0.1\%Eu^{3+}$ to 326 K for $5\%Eu^{3+}$. Morphological studies confirm that the synthesized powders consist of microsized grains (Figure 1f, see also Figure S32) with the uniformly distributed elements (Figure 1f).

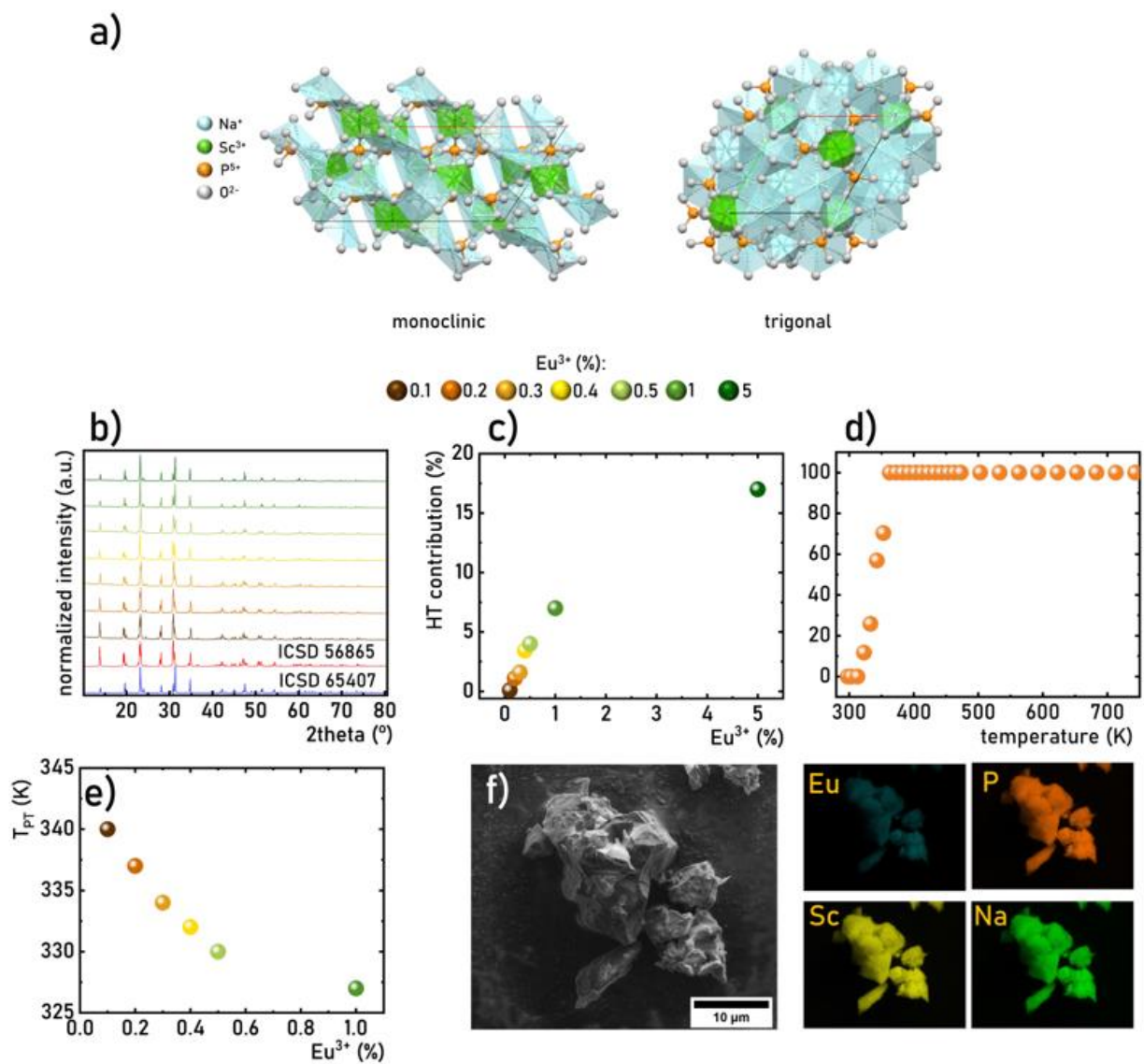

**Figure 1**. Visualization of the structure of the monoclinic and trigonal phases of $Na_3Sc_2(PO_4)_3$-a) the comparison of room temperature XRD patterns of $Na_3Sc_2(PO_4)_3$:$Eu^{3+}$ with different $Eu^{3+}$ concentration-b); the contribution of the trigonal phase of $Na_3Sc_2(PO_4)_3$ in the room temperature XRD patterns as a function of $Eu^{3+}$ concentration-c); the contribution of the trigonal phase of $Na_3Sc_2(PO_4)_3$ as a function of temperature for $Na_3Sc_2(PO_4)_3$:0.2%$Eu^{3+}$-d); temperature of the $\alpha \rightarrow \beta$ phase transition of $Na_3Sc_2(PO_4)_3$:$Eu^{3+}$ as a function of $Eu^{3+}$ concentration determined from the DSC analysis-e); representative SEM image -f) and corresponding elemental maps of the Eu (cyan), P (orange), Sc (yellow) and Na (green) for $Na_3Sc_2(PO_4)_3$:0.2%$Eu^{3+}$.

The luminescence properties of $Eu^{3+}$ ions are associated with the intraconfigurational *4f→4f* electronic transitions[51–54]. The shape of the emission spectra is primarily influenced by the radiative depopulation of the $^5D_0$ excited state to the $^4F_J$ multiplets. $Eu^{3+}$ ions stand out among lanthanides ions due to the high sensitivity of their luminescent properties to changes in the structural environment of the crystallographic sites they occupy within the host material[55–57]. This sensitivity manifests in several ways. Firstly, the $^5D_0 \rightarrow {}^7F_1$ electronic transition is a magnetic dipole transition, whereas the $^5D_0 \rightarrow {}^7F_2$ transition is of electric dipole character[57]. Unlike magnetic dipole, the intensity of electric dipole transitions depends on the point symmetry of the ion's environment and in many cases increases as the symmetry decreases. Therefore, the intensity ratio of the $^5D_0 \rightarrow {}^7F_2$ to $^5D_0 \rightarrow {}^7F_1$ transitions, often termed the asymmetry ratio, is sometimes used as a measure of point symmetry[58–61]. However, as noted by Binneman[58], the intensity of the $^5D_0 \rightarrow {}^7F_2$ band can be influenced by multiple factors, making it misleading to directly correlate an increase in asymmetry ratio with a decrease in point symmetry. Nonetheless, asymmetry ratio analysis remains a useful indicator of structural change. Furthermore, since the $^5D_0$ and $^7F_0$ levels do not split into Stark components, the $^5D_0 \rightarrow {}^7F_0$ electronic transition, observed in low-symmetry host materials, appears as a single emission line. The presence of multiple lines for this transition indicates emissions from non-equivalent crystallographic sites occupied by $Eu^{3+}$ ions. Additionally, the number of Stark components into which the $Eu^{3+}$ multiplets split depends on the host material's symmetry, with higher symmetry resulting in fewer Stark levels[58]. Therefore, for instance, in a monoclinic phase of $Na_3Sc_2(PO_4)_3:Eu^{3+}$, the $^7F_1$ and $^7F_2$ levels split into three and five Stark levels, respectively (Figure 2a), whereas, in trigonal symmetry, these multiplets split into two and three Stark levels, respectively. Due to these properties, $Eu^{3+}$ ions, often called 'optical structural probes'[58–62], are excellent candidates for analyzing structural changes in $Na_3Sc_2(PO_4)_3$. Emission spectra for $Na_3Sc_2(PO_4)_3:Eu^{3+}$, measured at three different temperatures (83 K, 373 K, 500 K) based on DSC analysis, represent $\alpha$, $\beta$ and $\gamma$ phases of $Na_3Sc_2(PO_4)_3$:$Eu^{3+}$, respectively (Figure 2b). In each analyzed phase, a single line corresponding to the $^5D_0 \rightarrow {}^7F_0$ transition is observed, indicating the emission corresponds to a single type of $Eu^{3+}$ luminescent center. Both the shape of the $Eu^{3+}$ emission bands and their relative intensities significantly depend on the crystallographic phase. For the monoclinic phase of $Na_3Sc_2(PO_4)_3$:$Eu^{3+}$, the emission bands display more Stark lines than in the trigonal phases (Figure 2b). Additionally, the intensity of the $^5D_0 \rightarrow {}^7F_2$ band increases in respect to the $^5D_0 \rightarrow {}^7F_1$ band as the structure changes from $\alpha$ to $\beta$. The obtained LIR parameter values for these phases are 1.62 for $\alpha$, 2.34 for $\beta$, and 2.3 for $\gamma$. Although less pronounced, differences between different $Na_3Sc_2(PO_4)_3$:$Eu^{3+}$ phases are also observed in the excitation spectra (Figure 2c, see also Figure S32). Regardless of the crystallographic phase, however, the excitation spectra are dominated by the $^7F_0 \rightarrow {}^5L_6$ band at approximately 393.5 nm, which was used as the excitation wavelength throughout this study. A comparison of luminescence decay profiles for all $Na_3Sc_2(PO_4)_3$:$Eu^{3+}$ phases indicates that the longest $\tau_{avr}$=3.37 ms was obtained in phase $\alpha$ (Figure 2d). The phase transition to $\beta$ shortened $\tau_{avr}$ to 2.47 ms, while the $\beta$-to-$\gamma$ transition did not alter the luminescence kinetics of $Eu^{3+}$ ions. Given that the $^5D_0$ emitting level is separated from the $^7F_6$ level by about 11,500 cm$^{-1}$, the probability of its nonradiative depopulation via multiphonon processes is low[63–67]. Thus, the observed $\tau_{avr}$ reduction can be attributed to an increased probability of radiative depopulation of the $^5D_0$ state associated with the structural phase transition.

These results clearly show that both the emission spectra and depopulation kinetics of the $^5D_0$ level of $Eu^{3+}$ ions are sensitive to thermally induced structural transitions in $Na_3Sc_2(PO_4)_3$:$Eu^{3+}$. As demonstrated by DSC studies, increasing the $Eu^{3+}$ ion concentration lowers the structural phase transition temperature in $Na_3Sc_2(PO_4)_3$:$Eu^{3+}$, consequently impacting the shape of the obtained room temperature emission spectra. As the $Eu^{3+}$ concentration increases, the intensity of the

$^5D_0 \rightarrow {}^7F_2$ band rises in respect to the $^5D_0 \rightarrow {}^7F_1$ band (Figure 2e). The ratio of these intensities increases monotonically to a value of 2.59 for 1%Eu$^{3+}$, beyond which further doping does not alter this parameter (Figure 2f). This enhancement in the luminescence intensity ratio (LIR) for low Eu$^{3+}$ concentrations is attributed to an increase in the high-temperature phase of Na$_3$Sc$_2$(PO$_4$)$_3$:Eu$^{3+}$ and thus an increased presence of Eu$^{3+}$ ions in the trigonal phase. However, LIR values at Eu$^{3+}$ concentrations above 1% exceed those recorded for Na$_3$Sc$_2$(PO$_4$)$_3$:0.2%Eu$^{3+}$, suggesting that effects beyond phase transition influence the material at high dopant concentrations. These additional effects likely stem from additional symmetry changes due to the reduction of the elemental cell of Na$_3$Sc$_2$(PO$_4$)$_3$:Eu$^{3+}$ associated with the ionic radius difference between the Eu$^{3+}$ dopant and the Sc$^{3+}$ cation it replaces[68].

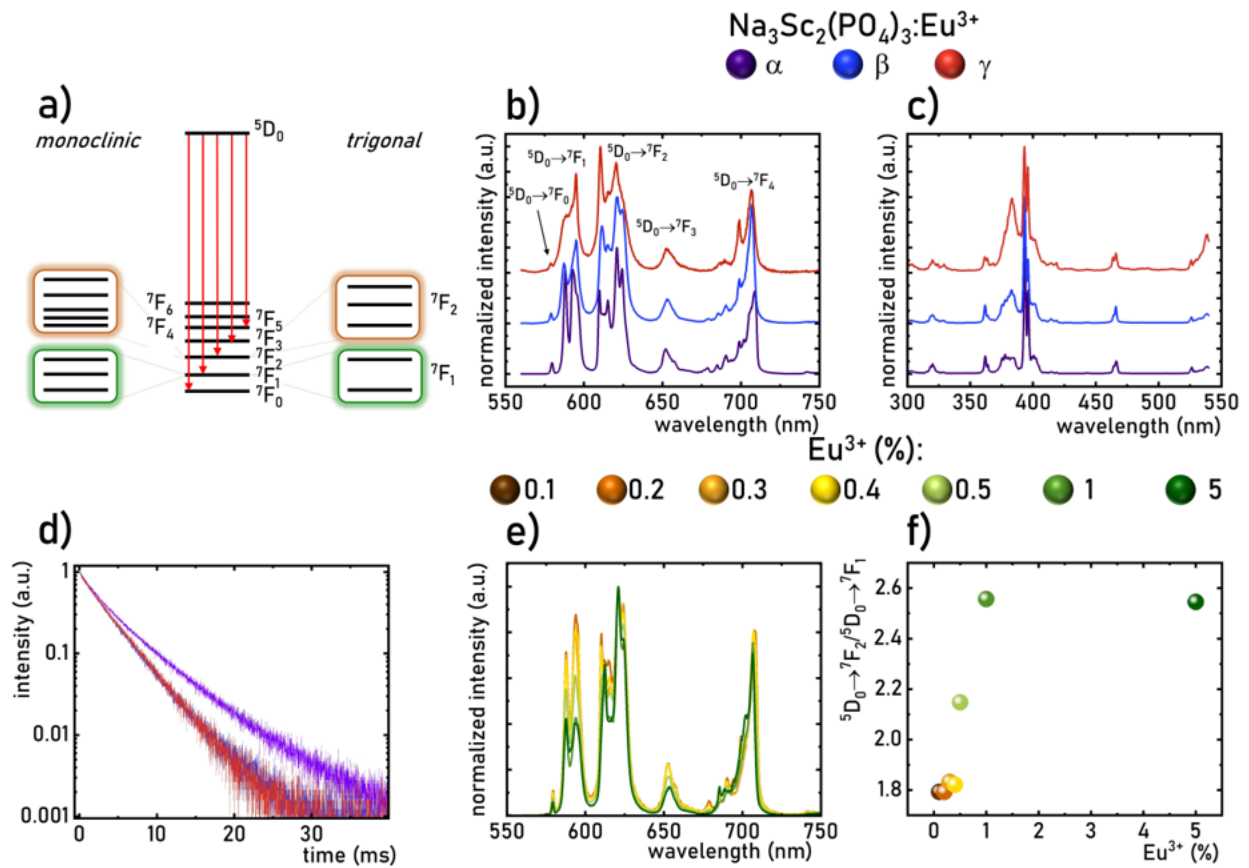

**Figure 2**. Simplified energy level diagram of $Eu^{3+}$ ions-a); comparison of the emission spectra of $Na_3Sc_2(PO_4)_3$: 0.2%$Eu^{3+}$ measured at 83 K (violet line) 363 K (blue line) and 603 K (red line)-b) and corresponding excitation spectra ($\lambda_{exc}$=393.5 nm)-c); and luminescence decay profiles ($\lambda_{exc}$=393.5 nm, $\lambda_{em}$=620 nm)-d); the comparison of the normalized room temperature emission spectra of $Na_3Sc_2(PO_4)_3$: $Eu^{3+}$ with different concentration of dopant ions-e) and the asymmetric ratio as a function of $Eu^{3+}$ concentration-f).

As demonstrated above, the structural phase transition in $Na_3Sc_2(PO_4)_3$:$Eu^{3+}$ significantly affects the shape of the emission spectra of $Eu^{3+}$ ions (see also Figures S33-38). More pronounced changes are expected during the $\alpha$–$\beta$ transition due to the significant symmetry alteration in the structure. The analysis of the emission spectra of $Na_3Sc_2(PO_4)_3$:$Eu^{3+}$ measured as a function of temperature reveals clear and gradual changes in spectral shape (Figure 3a). Increasing the temperature results in a progressive reduction in the intensity of Stark lines associated with the luminescence of $Eu^{3+}$ ions in the monoclinic phase, accompanied by an enhancement in those associated with the trigonal phase. These transitions alter the shape of each emission band as well as their relative intensities. To quantify these changes, four different luminescence intensity ratio (*LIR*) parameters were proposed as follows:

$$LIR_1 = \frac{\int_{583nm}^{597nm} \left(^5D_0 \to {}^7F_1\right) d\lambda}{\int_{606nm}^{631nm} \left(^5D_0 \to {}^7F_2\right) d\lambda} \qquad (3)$$

$$LIR_2 = \frac{\int_{693nm}^{695nm} \left(^5D_0 \to {}^7F_4\right) d\lambda}{\int_{691nm}^{693nm} \left(^5D_0 \to {}^7F_4\right) d\lambda} \qquad (4)$$

$$LIR_3 = \frac{\int_{611nm}^{613nm} \left(^5D_0 \to {}^7F_2\right) d\lambda}{\int_{608nm}^{610nm} \left(^5D_0 \to {}^7F_2\right) d\lambda} \qquad (5)$$

$$LIR_4 = \frac{\int_{697nm}^{702nm} \left(^5D_0 \to {}^7F_4\right) d\lambda}{\int_{690nm}^{697nm} \left(^5D_0 \to {}^7F_4\right) d\lambda} \qquad (6)$$

A comparison of the thermal dependence of $LIR_i$ for a representative sample of Na$_3$Sc$_2$(PO$_4$)$_3$:0.2%Eu$^{3+}$, shown in Figure 3b, indicates significant changes in each $LIR$ parameter with increasing temperature near the structural phase transition (see also Figures S39-42). However, the most pronounced variations are observed for $LIR_3$, and thus the subsequent analysis in this work focuses on this thermometric parameter. To investigate the effect of Eu$^{3+}$ ion concentration on the thermometric performance of the ratiometric luminescence thermometer of Na$_3$Sc$_2$(PO$_4$)$_3$:Eu$^{3+}$, the thermal dependences of $LIR_3$ are presented in a normalized form (normalized to its initial value obtained at 83 K) (Figure 3c). This analysis reveals several noteworthy trends. For Eu$^{3+}$ ion concentrations below 0.5%, $LIR_3$ initially increases slightly with rising temperature up to approximately 300 K, above which exhibits a rapid increase, peaking at around 370 K. Beyond this temperature, a further increase results in a gradual decline in $LIR_3$. The threshold temperature at which a sharp rise in $LIR_3$ is observed decreases with increasing Eu$^{3+}$ ion concentration, consistent with the lowering of the phase transition temperature. For a concentration of 0.3%Eu$^{3+}$, a nearly twofold increase in $LIR_3$ values is observed in the 300-380 K thermal range. Conversely, for Eu$^{3+}$ concentrations exceeding 1%, $LIR_3$ exhibits a monotonic decrease across the entire temperature range analyzed. It is critical to note that for reliable temperature sensing with a luminescent thermometer, the temperature-dependent parameter must exhibit a monotonic trend within the operational thermal range. Consequently, the decrease in $LIR_3$ above approximately 450 K renders $Na_3Sc_2(PO_4)_3$:$Eu^{3+}$ unsuitable for temperature sensing beyond this threshold. Furthermore, the thermal operating range gradually narrows with increasing $Eu^{3+}$ ion concentration. Quantitative evaluation of the observed changes in $LIR_3$ is achieved by determining the relative sensitivity ($S_R$) using the following equation:

$$S_R = \frac{1}{LIR}\frac{\Delta LIR}{\Delta T} \cdot 100\% \qquad (7)$$

where $\Delta LIR$ represents the change in $LIR$ corresponding to the change in temperature by $\Delta T$. The $S_R$ values for $Na_3Sc_2(PO_4)_3$:$Eu^{3+}$ gradually increase with temperature, reaching a maximum at approximately 390 K (Figure 3d). However, the temperature at which $S_{Rmax}$ is observed decreases progressively with higher $Eu^{3+}$ ion concentrations. The maximum $S_R$ value increases with rising dopant ion concentration, peaking at $S_{Rmax}$=3.4% $K^{-1}$ for 0.3%$Eu^{3+}$, after which further increases in $Eu^{3+}$ concentration result in a decline in $S_R$ (Figure 3e). Notably, all the $S_R$ values obtained exceed 1.7% $K^{-1}$, a relatively high sensitivity compared to other ratiometric luminescence thermometers based on lanthanide ion emissions[6,16,69].

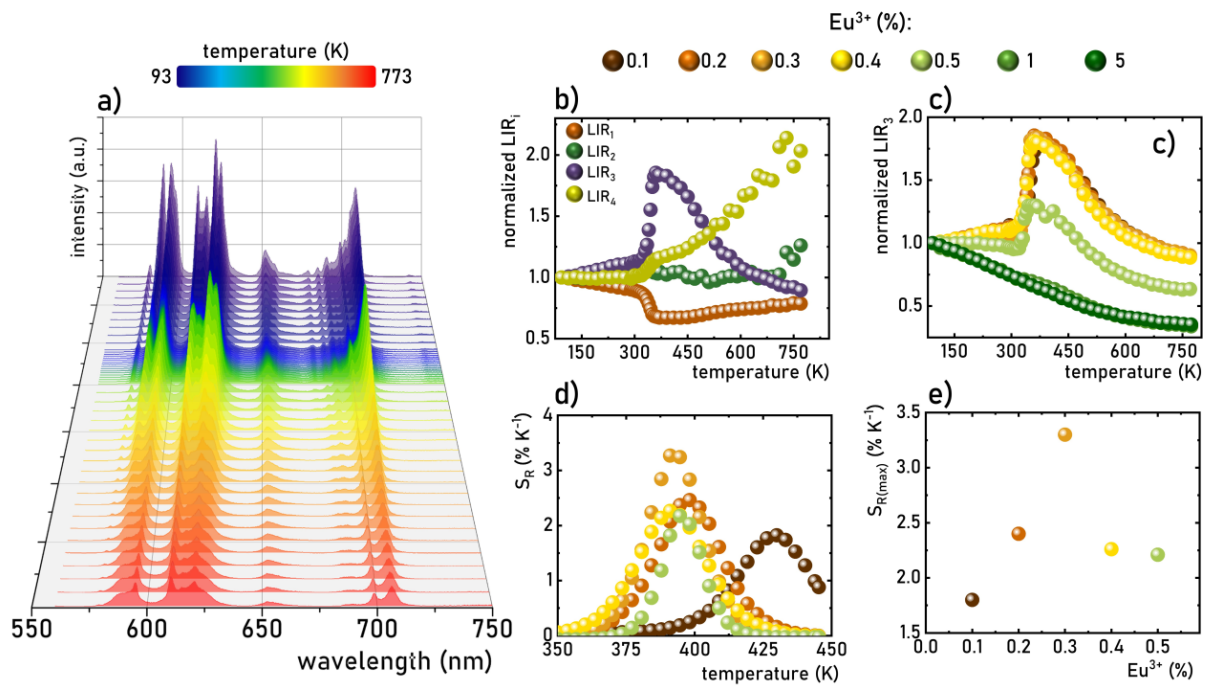

**Figure 3.** Luminescence spectra of $Na_3Sc_2(PO_4)_3$:0.2%$Eu^{3+}$ measured as a function of temperature -a); the thermal dependence of normalized $LIR_i$ for $Na_3Sc_2(PO_4)_3$:0.2%$Eu^{3+}$ -b); the thermal dependence of normalized $LIR_3$ for different $Eu^{3+}$ ions concentrations-c) and corresponding thermal dependence of $S_R$ -d); the influence of the $Eu^{3+}$ ions concentrations on the $S_{Rmax}$ of the ratiometric luminescence thermometers based on $Na_3Sc_2(PO_4)_3$:0.2%$Eu^{3+}$ -e).

As discussed above, structural changes associated with the thermally induced phase transition in $Na_3Sc_2(PO_4)_3$:$Eu^{3+}$ result in notable variations in luminescence kinetics of $Eu^{3+}$ ions (Figures S43-48). The analysis of luminescence decay profiles of $Eu^{3+}$ ions in $Na_3Sc_2(PO_4)_3$:$Eu^{3+}$ yields several intriguing observations. As the temperature increases beyond 83 K, initially minor changes in the lifetimes are observed (Figure 4a). However, beyond a certain threshold temperature, a pronounced shortening of the $\tau_{avr}$ becomes evident. This trend is clearly reflected in the thermal dependence of $\tau_{avr}$ (Figure 4b). For $Eu^{3+}$ ion concentrations below 0.4%, a $\tau_{avr}$ of 3.30 ms is observed, and increasing the temperature does not significantly affect its value until approximately 330 K, corresponding to the $\alpha \rightarrow \beta$ phase transition. In the temperature range associated with this phase transition, a sharp reduction in $\tau_{avr}$ to approximately 2.5 ms is observed, above which further increases in temperature do not result in significant changes. For a concentration of 0.5% $Eu^{3+}$, a similar trend is noted, though the initial lifetime value at 83 K is lower ($\tau_{avr}$ =2.81 ms), and further thermal shortening of the lifetime becomes apparent above approximately 550 K. At higher dopant concentrations, the $\tau_{avr}$ exhibit minimal thermal variation until about 500 K, beyond which its thermal shortening is observed. It is worth to underline that for 0.1-0.4%$Eu^{3+}$ dopant concentrations, the temperature at which $\tau_{avr}$ begins to decrease shifts to lower values as the concentration of $Eu^{3+}$ increases, reflecting a corresponding reduction in the phase transition temperature. A comparison of $\tau_{avr}$ values at 83 K indicates that below 0.4%$Eu^{3+}$, $\tau_{avr}$ remains independent of dopant ion concentration (Figure 4c). This confirms the absence of interionic interactions between dopants, consistent with the energy level diagram of $Eu^{3+}$ ions. The reduction in $\tau_{avr}$ observed for 0.5%$Eu^{3+}$ may suggest additional structural changes related to the difference in ionic radii between the $Eu^{3+}$ dopant ions and the host cations, as mentioned earlier. Interestingly, the $\tau_{avr}$ values after the $\alpha \rightarrow \beta$ phase transition are nearly independent of dopant ion concentration.

The relative sensitivity for a lifetime-based luminescence thermometer can be calculated analogously to Equation 7 (substituting *LIR* with $\tau_{avr}$). The maximum $S_R$ values are observed at temperatures corresponding to the structural phase transition (Figure 4d). However, the absence of $\tau_{avr}$ shortening for $Eu^{3+}$ ion concentrations of 1% and 5% near 340 K results in negligible $S_R$ values in this range. The highest $S_R$ value was recorded for 0.3%$Eu^{3+}$, with $S_R$=1.0% $K^{-1}$, and this value decreases as the $Eu^{3+}$ ion concentration changes (Figure 4e). Notably, a linear correlation was observed between the temperature at which $S_{Rmax}$ occurs and the increasing dopant ion concentration (Figure 4f). This result provides direct confirmation of the potential for fine-tuning the thermometric performance of the luminescent thermometer by adjusting the dopant ion concentration. Such tunability is highly significant in meeting the requirements of various potential applications.

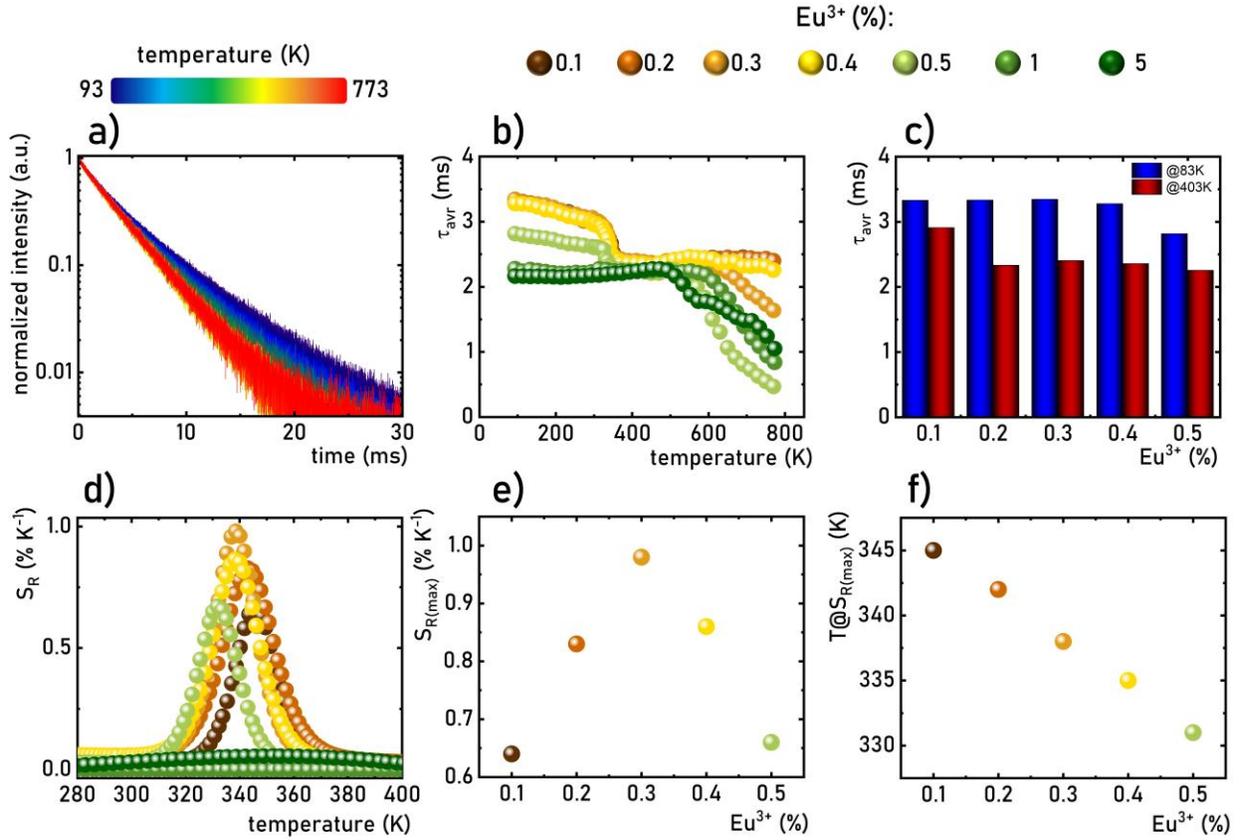

**Figure 4**. Luminescence decay profile of $Na_3Sc_2(PO_4)_3$:0.2%$Eu^{3+}$ measured as a function of temperature - a); $\tau_{avr}$ as a function of temperature for different concentrations of $Eu^{3+}$ ions-b); comparison of the $\tau_{avr}$ at 83 K and 403 K for different concentrations of $Eu^{3+}$ ions-c); thermal dependence of $S_R$ of lifetime-based luminescence thermometers on $Na_3Sc_2(PO_4)_3$:$Eu^{3+}$ with different concentration of $Eu^{3+}$ ions-d) $S_{Rmax}$ as a function of dopant concentration-e) and influence of $Eu^{3+}$ concentration on a temperature at which $S_{Rmax}$ was obtained-f).

As shown above, the changes induced by the structural phase transition in the local environment of $Eu^{3+}$ ions in $Na_3Sc_2(PO_4)_3$:$Eu^{3+}$ result in a change in the intensity ratio of the $^5D_0 \rightarrow {}^7F_2$ to $^5D_0 \rightarrow {}^7F_1$ emission bands. Since the former band falls within the red region of the spectrum and the latter within the yellow region, relative changes in their intensities lead to a variation in the colour of the emitted light. However, as shown in the CIE 1931 chromaticity diagram (Figure 5a), this change in color is relatively small. A detailed analysis of the chromaticity coordinates of the emitted light from $Na_3Sc_2(PO_4)_3$:0.2%$Eu^{3+}$ as a function of temperature reveals that, near the phase transition temperature, $x$ increases from 0.6416 to 0.6493, while $y$ decreases from 0.35828 to 0.35032 (Figure 5b). These shifts enable temperature readout based on the emitted light's color. Nevertheless, the relative sensitivities calculated from the chromatic coordinates yield modest values (for $x$, $S_{Rmax}$=0.036 % $K^{-1}$; for $y$, $S_{Rmax}$=-0.064457 % $K^{-1}$) (Figure 5c). The negative values of the $S_R$ in the case of $y$ coordinate results from thermal reduction of $y$ value. On the other hand, a comparison of the $Eu^{3+}$ ion emission spectra for the low-temperature (LT) and high-temperature (HT) phases of $Na_3Sc_2(PO_4)_3$:$Eu^{3+}$ with the spectral sensitivity ranges of RGB channels of a conventional digital camera reveals that while the R channel captures intensity contributions from both the $^5D_0 \rightarrow {}^7F_2$ and $^5D_0 \rightarrow {}^7F_1$ bands, the G channel is primarily sensitive to the $^5D_0 \rightarrow {}^7F_1$ (Figure 5d). When an image is captured using a digital camera, the sensor simultaneously collects intensity data from all RGB channels. Based on this principle, we propose using the phase transition-induced changes in the emission spectra of $Na_3Sc_2(PO_4)_3$:0.2%$Eu^{3+}$ for temperature measurement and two-dimensional imaging. To demonstrate this, a series of photographs of $Na_3Sc_2(PO_4)_3$:0.2%$Eu^{3+}$ powder placed on a temperature-controlled heating-cooling stage under continuous optical excitation were taken (Figure 5e, Figure S49). From the captured images, intensity maps for the R and G channels were extracted and analyzed (Figure 5f). Due to the limited sensitivity of the G channel in the spectral range corresponding to the $^5D_0 \rightarrow {}^7F_1$ transition, the G channel displayed relatively low intensity values (intensity increases 20 times in Figure 5f, see also Figure S50). These intensity maps were subsequently divided to calculate the R/G intensity ratio (Figure S51). As expected, the R/G intensity ratio decreased with increasing temperature, directly validating the effectiveness of the proposed approach.

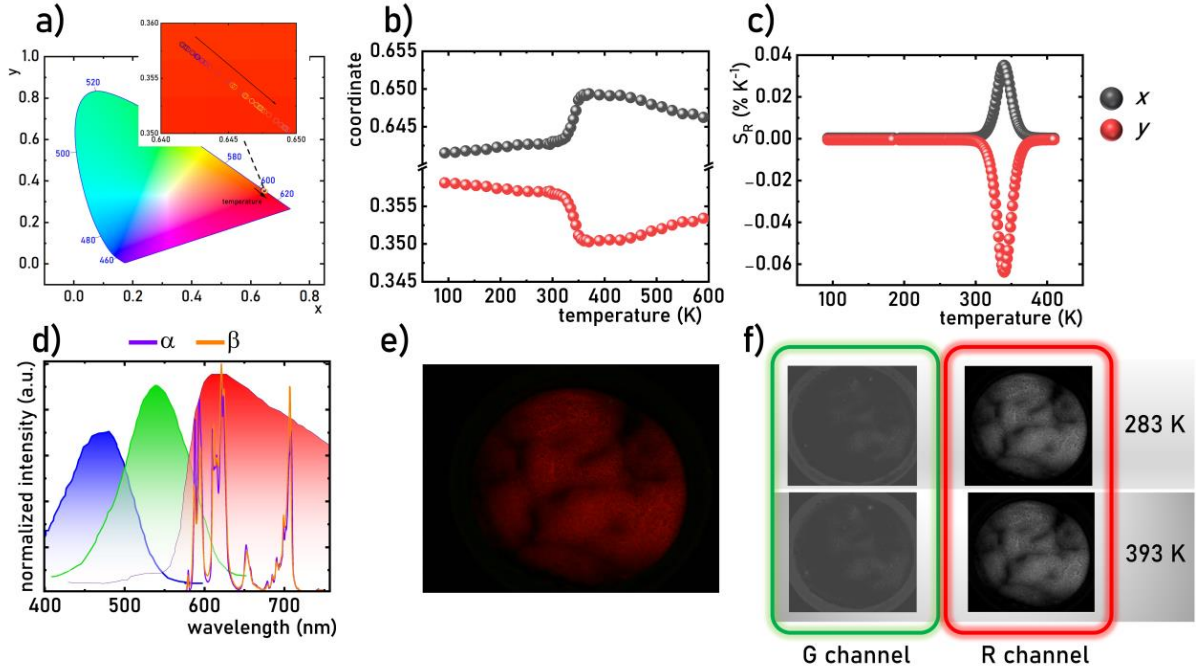

**Figure 5**. The influence of temperature on the CIE 1931 chromatic coordinates for $Na_3Sc_2(PO_4)_3:0.2\%Eu^{3+}$ -a); thermal dependence of $x$ (black dots) and $y$ (red dots) chromatic coordinates -b) and corresponding relative sensitivities -c); the comparison of normalized emission spectra of low temperature (violet line) and high temperature (orange line) phase of the $Na_3Sc_2(PO_4)_3:0.2\%Eu^{3+}$ with the spectra of sensitivities of RGB channels of digital camera-d); representative image of luminescence of $Na_3Sc_2(PO_4)_3:0.2\%Eu^{3+}$ measured at 283 K-e) and the two-dimensional maps of luminescence intensities registered in G and R channels for $Na_3Sc_2(PO_4)_3:0.2\%Eu^{3+}$ at 283K and 393K-f).

Given that the structural phase transition in $Na_3Sc_2(PO_4)_3$:0.2%$Eu^{3+}$ induces a modification in the ratio of the emission intensities collected in the red and green channels of the digital camera, this phenomenon can be effectively exploited not only for temperature spot-readout but also for spatially resolved temperature imaging. The uniqueness of this effect lies in the fact that, under typical conditions, each of the most intense emission bands of $Eu^{3+}$ ions ($^5D_0 \rightarrow {}^7F_1$ and $^5D_0 \rightarrow {}^7F_2$) originates from the same emitting level ($^5D_0$). Therefore, significant temperature-dependent changes in the intensity ratio of these bands are generally not expected. The observation of such a dependence in the case of $Na_3Sc_2(PO_4)_3$:0.2%$Eu^{3+}$ is thus particularly noteworthy. To evaluate the potential of this phenomenon for temperature imaging, an experimental setup was designed in which the $Na_3Sc_2(PO_4)_3$ phosphor powder was spread on a quartz substrate (Figure 6 a-b), which was then placed on a heating stage maintained at a constant temperature of 453 K. To inhibit direct heat transfer and induce a thermal gradient, the quartz holder was thermally insulated from the heating stage using steel rings with a diameter of 5 mm, as illustrated in Figure 6b. After confirming thermal stabilization of the heating stage using thermovision camera (Figure 6c), the sample was positioned on the stage and time-sequenced digital images were captured at intervals of 1.33 s. From each captured image, the red and green channels were extracted, and their pixel-wise ratio (R/G) was calculated to generate two-dimensional R/G distribution maps (Figure 6 d-f). Using a previously established calibration curve, obtained under controlled isothermal conditions (Figure S51), the R/G ratio maps were converted into corresponding temperature maps. The results, presented in Figure 6h-k (see also Figure S52), clearly illustrate the development of a temperature gradient across the $Na_3Sc_2(PO_4)_3$:0.2%$Eu^{3+}$ sample, with an observable increase in maximum temperature over time. The temporal evolution of temperature derived from the luminescence maps shows strong agreement with values obtained from the thermal imaging camera, with a maximum deviation of just 0.5 K (Figure 6l, Figure S53). Notably, the luminescence-based thermal maps achieved a spatial resolution of 0.2 mm and allowed for detailed temperature profiling along arbitrary cross-sections of the sample (Figure 6 m). While the temperature accuracy of the luminescence-based measurements is in the same order of magnitude to that of conventional thermovision camera, a significant advantage of the luminescence thermometry approach is its ability to image temperature through transparent media in the visible spectral range. This was demonstrated by placing a quartz lid with an optical window over the sample (Figure S54). In this configuration, the thermal imaging camera could only detect the surface temperature of the quartz window, whereas luminescence thermometry successfully captured the underlying temperature gradient of the phosphor. Moreover, luminescence thermometry offers spatial specificity: the temperature information is derived exclusively from the region containing the luminescent material, allowing for selective temperature mapping of the target area. This stands in contrast to conventional infrared thermal imaging, which measures the temperature of any object within its optical path, potentially reducing specificity in complex or multilayered systems.

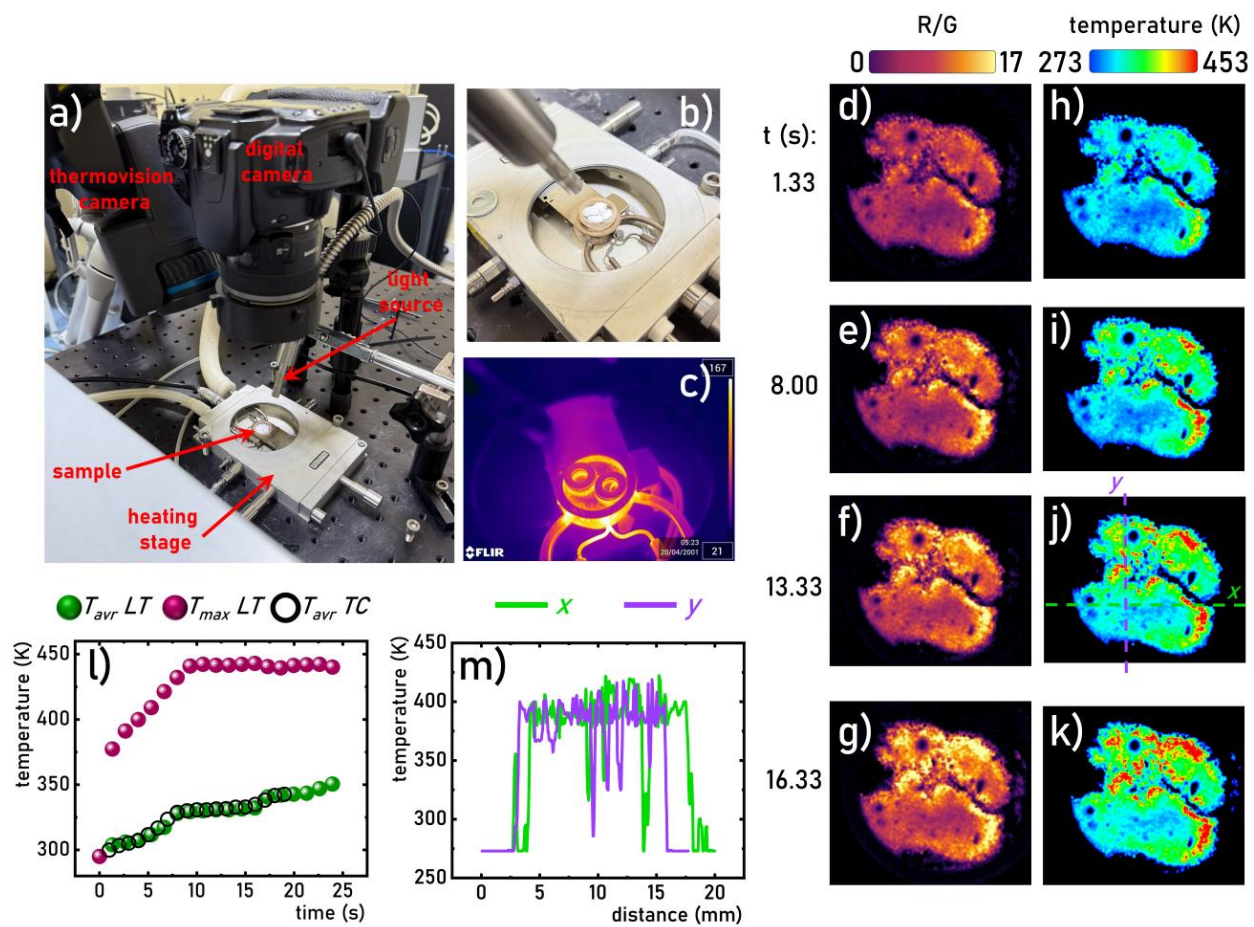

**Figure 6**. Photo of the experimental setup used for thermal imaging based on the optical response of the Na$_3$Sc$_2$(PO$_4$)$_3$:0.2%Eu$^{3+}$ -a); Na$_3$Sc$_2$(PO$_4$)$_3$:0.2%Eu$^{3+}$ sample placed on the heating stage-b) photo of the heating elements recorded using thermovision camera -c); 2-dimensional maps of the R/G ratio obtained from the digital camera recorded as a function of time d-g); and corresponding thermal maps - h-k); temporal dependence of the maximal ($T_{max}$) and average ($T_{avr}$) temperature of the phosphor determined using Na$_3$Sc$_2$(PO$_4$)$_3$:0.2%Eu$^{3+}$ luminescence thermometer (*LT*) and thermovision camera (*TC*) -l); temperature determined using luminescence thermometry across *x* and *y* axis marked in Figure 6j) -m).

The topic of luminescent thermometry based on structural phase transitions has recently gained growing interest, as evidenced by an increasing number of publications exploring this approach (Table 1)[31–34,70–75]. However, the inherently localized nature of first-order structural phase transitions limits the structural changes, and thus the associated spectroscopic modifications, to a relatively narrow temperature range. As a result, the temperature at which the maximum relative sensitivity is observed often defines the effective operating range of the luminescent thermometer. This limitation arises from the fact that, in most phase-transition-based systems, a monotonic change in the luminescence intensity ratio (*LIR*) is typically confined to a window of approximately ±50 K around the phase transition temperature. Therefore, although such systems can exhibit exceptionally high $S_R$ values, tailoring the thermometer to a specific application-defined temperature range requires identifying host materials with appropriately tuned phase transition temperatures. For instance, while LiYO$_2$:Ln$^{3+}$ materials are known for their high relative sensitivities, their operational ranges are generally centered near room temperature[31–33,70,72,74,75]. To overcome this constraint, new host matrices such as LaGaO$_3$[70] and Na$_3$Sc$_2$(PO$_4$)$_2$ [71] have recently been proposed. Notably, in contrast to Na$_3$Sc$_2$(PO$_4$)$_2$:Yb$^{3+}$, the Na$_3$Sc$_2$(PO$_4$)$_2$:Eu$^{3+}$ system described in this work exhibits significantly higher relative sensitivity. Furthermore, this study presents, for the first time, the feasibility of utilizing a structural phase transition for two-dimensional thermal imaging. This result not only underscores the practical potential of phase-transition-based luminescent thermometry but also demonstrates, crucially and for the first time, that thermometric systems relying solely on Eu$^{3+}$ luminescence can be effectively applied for thermal imaging applications.

**Table 1**. Comparison of the thermometric performance of ratiometric luminescence thermometers based on the first-order structural phase transition

| Thermometer | LT phase | HT phase | LIR | $S_{Rmax}$ [% K$^{-1}$] | $T@S_{Rmax}$ [K] | Ref |
|---|---|---|---|---|---|---|
| LiYO$_2$:5%Yb$^{3+}$ | monoclinic | tetragonal | $^2F_{5/2}\rightarrow{}^2F_{7/2}$ / $^2F_{5/2}\rightarrow{}^2F_{7/2}$ | 5.3 | 280 | [73] |
| $Na_3Sc_2(PO_4)_3:Yb^{3+}$ | monoclinic | trigonal | $^2F_{5/2}\rightarrow{}^2F_{7/2}$ / $^2F_{5/2}\rightarrow{}^2F_{7/2}$ | 1.5 | 340 | [71] |
| $LiYO_2:Pr^{3+}$ | monoclinic | tetragonal | $^3P_0\rightarrow{}^3H_4$ / $^1D_2\rightarrow{}^3H_4$ | 23.04 | 329 | [75] |
| $LiYO_2:1\%Eu^{3+}$ | monoclinic | tetragonal | $^5D_0\rightarrow{}^7F_2$ / $^5D_0\rightarrow{}^7F_2$ | 12.5 | 305 | [34] |
| $LiYO_2:1\%Eu^{3+},30\%Yb^{3+}$ | monoclinic | tetragonal | $^5D_0\rightarrow{}^7F_2$ / $^5D_0\rightarrow{}^7F_2$ | 2.1 | 180 | [32] |
| $LiYO_2:1\%Eu^{3+},40\%Gd^{3+}$ | monoclinic | tetragonal | $^5D_0\rightarrow{}^7F_2$ / $^5D_0\rightarrow{}^7F_2$ | 1.4 | 550 | [32] |
| $LiYO_2:0.1\%Nd^{3+}$, | monoclinic | tetragonal | $^4F_{3/2}\rightarrow{}^4I_{9/2}$ / $^4F_{3/2}\rightarrow{}^4I_{9/2}$ | 7.9 | 291 | [72] |
| $LiYO_2: 1\%Er^{3+}, 10\%Yb^{3+}$ | monoclinic | tetragonal | $^4S_{3/2}\rightarrow{}^4I_{15/2}$ /$^4S_{3/2}\rightarrow{}^4I_{15/2}$ | 2.5 | 240 | [31] |
| $LiYO_2: Dy^{3+}$, | monoclinic | tetragonal | $^4F_{9/2}\rightarrow{}^6H_{13/2}$ / $^4F_{9/2}\rightarrow{}^6H_{13/2}$ | 35.24 | 310 | [76] |
| $LaGaO_3:0.1\%Eu^{3+}$ | orthorhombic | trigonal | $^5D_0\rightarrow{}^7F_2$ / $^5D_0\rightarrow{}^7F_2$ | 5.4 | 456 | [70] |
| $Na_3Sc_2(PO_4)_3:Eu^{3+}$ | monoclinic | trigonal | $^5D_0\rightarrow{}^7F_2$ / $^5D_0\rightarrow{}^7F_2$ | 3.4 | 390 | This work |

**Conclusions**

In this study, the spectroscopic properties of $Na_3Sc_2(PO_4)_3:Eu^{3+}$ were investigated as a function of temperature and $Eu^{3+}$ ion concentration to develop a novel luminescent thermometer based on a structural phase transition. The results show that an increase in temperature above approximately 350 K induces a phase transition in $Na_3Sc_2(PO_4)_3:Eu^{3+}$ from a monoclinic to a trigonal structure. This transition significantly alters the point symmetry of the crystallographic site occupied by $Eu^{3+}$ ions, thereby modifying their spectroscopic properties. The studies revealed that the monoclinic-to-trigonal phase transition reduces the number of Stark levels into which the $^4F_J$ multiplets split, changes the luminescence intensity ratio of the $^5D_0\rightarrow{}^7F_2$ to $^5D_0\rightarrow{}^7F_1$ transitions of $Eu^{3+}$ ions, and shortens the lifetime of the $^5D_0$ level. These changes enabled the development of a luminescence thermometer using both ratiometric and lifetime-based approaches. The highest sensitivities, 3.4% $K^{-1}$ and 1.0% $K^{-1}$, were achieved for ratiometric and lifetime-based luminescence thermometers, respectively, in $Na_3Sc_2(PO_4)_3$:0.3%$Eu^{3+}$. Additionally, it was shown that increasing the $Eu^{3+}$ ion concentration to 0.5% resulted in a reduction of the phase transition temperature, allowing a corresponding decrease in the temperature at which $S_{Rmax}$ was observed. However, for $Eu^{3+}$ concentrations above 1%, the monoclinic-to-trigonal phase transition was not observed, likely due to the stabilization of the trigonal phase of $Na_3Sc_2(PO_4)_3$:$Eu^{3+}$ caused by the ionic radius mismatch between the host material and dopant cations. The developed thermometers exhibited relatively high relative sensitivity and a thermal operating range of approximately 75 K. Moreover, for the first time, the applicative potential of first order phase transition based luminescence thermometers in thermal imaging was demonstrated here. These findings demonstrate that the synergy between the high sensitivity of $Eu^{3+}$ spectroscopic properties and thermally induced structural phase transitions can be effectively utilized for temperature sensing and imaging.

## Acknowledgements

This work was supported by National Science Center (NCN) Poland under project no. DEC-UMO-2022/45/B/ST5/01629. The authors would like to express their gratitude to prof. Artur Bednarkiewicz for his help in processing the luminescence maps.